\documentstyle[prl,aps,graphicx,floats]{revtex}
\begin{document}
\draft

\title{Coalescence of the Fermi-surface-related diffuse intensity peaks 
in disordered alloys}
\author{Igor Tsatskis$^{\dag \ddag}$}
\address{Department of Earth Sciences, University of Cambridge,
Downing Street, Cambridge CB2 3EQ, United Kingdom}
\maketitle

\begin{abstract}
The possibility of disappearance of the diffuse-intensity peak splitting induced by 
the Fermi surface (i.e., of coalescence of the intensity maxima) with decreasing 
temperature is predicted. The underlying mechanism is the compensation of the 
reciprocal-space curvatures of the self-energy and the interaction. The theory also 
describes similar results obtained earlier for two low-dimensional models with 
competing interactions. The coalescence is compared with the recently observed 
``thermal'' splitting in Pt-V which can be explained in the same way.
\end{abstract}

\begin{figure}
\begin{center}
\includegraphics[angle=0]{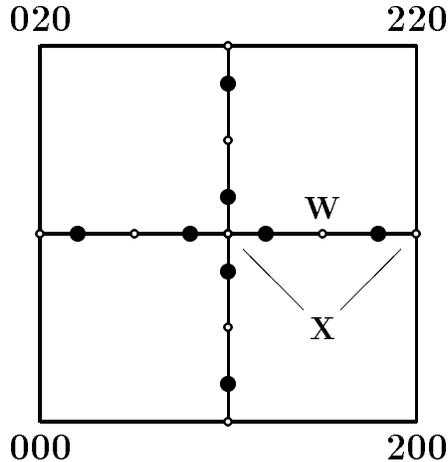}
\end{center}
\caption{The reciprocal-space geometry of the splitting. The (hk0) plane is shown. 
Filled circles mark the positions of the SRO diffuse-intensity peaks. Small open circles 
correspond to the SPs $X$ and $W$.}
\label{f1}
\end{figure}

In the recent paper (Tsatskis 1998a) we proposed an explanation of the temperature 
dependence of the diffuse-intensity peak splitting found for the disordered Cu$_{3}$Au 
alloy (Reichert, Moss and Liang 1996; see also Reichert, Tsatskis and Moss 1997). 
The splitting is associated with the Fermi surface (FS) effects and is observed also 
for other FCC alloy systems: Cu-Al (Scattergood, Moss and Bever 1970, Sch\"{o}nfeld 
{\em et al.} 1996), Cu-Pd (Ohshima and Watanabe 1973, Ohshima, Watanabe and Harada 1976, 
Saha, Koga and Ohshima 1992), Cu-Pt (Ohshima and Watanabe 1973, Saha and Ohshima 1993), 
etc. It is the consequence of the indirect interaction between alloy atoms via 
conduction electrons in those cases where the FS has flat or nested (i.e., identically 
curved) areas (e.g., Krivoglaz 1996). The resulting effective long-range pairwise 
interatomic interaction is characterized by minima at some positions between the 
$X=(110)$ and $W=(1 \frac{1}{2} 0)$ special (Lifshitz) points (SPs); this feature is 
then reflected in the short-range order (SRO) diffuse intensity (Fig.~\ref{f1}). 
According to the proposed theory, the temperature-dependent splitting is the 
non-mean-field effect and the result of the wavevector dependence of the so-called 
self-energy. The self-energy $\Sigma$ of the pair correlation function (for detailed 
discussion and relation to thermodynamics, see Tsatskis 1998b) enters the expression 
for the SRO intensity,
\begin{equation}
I({\bf k}) = \frac{1}{ c(1-c) \left[ - \Sigma({\bf k}) +
2 \beta V({\bf k}) \right] } \ , \label{1}
\end{equation}
where ${\bf k}$ is the wavevector, $I({\bf k})$ is the intensity in Laue units, 
$c$ is the concentration, $\beta=1/T$, $T$ is the temperature in energy units 
and $V({\bf k})$ is the Fourier transform of the pair ordering potential 
$V_{ij}=(V^{AA}_{ij}+V^{BB}_{ij})/2-V^{AB}_{ij}$. Potential $V^{\alpha \beta}_{ij}$ 
corresponds to the interaction between an atom of type $\alpha$ at site $i$ and 
an atom of type $\beta$ at site $j$. 

The ${\bf k}$-dependence of the self-energy leads to the temperature-dependent 
shift of the $I({\bf k})$ peak with respect to the corresponding minimum of the 
interaction $V({\bf k})$ and, as a result, to the temperature dependence of the 
splitting. The positions of the $I({\bf k})$ extrema are determined by the condition 
$\partial_{k} I = 0$ which gives
\begin{equation}
2 \, \partial_{k} V = T \, \partial_{k} \Sigma \ . \label{2}
\end{equation}
Here $k$ is the wavenumber along a line (e.g., (h10) or (1k0) in Fig.~\ref{f1}) 
containing off-SP peaks; in what follows only the $I({\bf k})$ profile along this line 
is analysed. Eq.~(\ref{2}) shows that the ${\bf k}$-dependent self-energy shifts the 
intensity peak position away from the $V({\bf k})$ minimum. The right-hand side of 
Eq.~(\ref{2}) contains $T$ as a factor, and $\Sigma({\bf k})$ itself is a function of 
$T$, while the left-hand side is $T$-independent. The shift and the splitting (which 
are linearly related) depend therefore on temperature. In the two widely used theories 
of SRO, the mean-field Krivoglaz-Clapp-Moss (KCM) approximation (Krivoglaz 1969, Clapp 
and Moss 1966, 1968) and the spherical model (SM) (e.g., Brout 1965), the self-energy 
is ${\bf k}$-independent, the right-hand side of Eq.~(\ref{2}) vanishes, the $I({\bf k})$ 
peaks coincide with the minima of $V({\bf k})$, and the splitting does not change with 
temperature. 

This treatment, however, implicitly assumes that in Eq.~(\ref{1}) the 
${\bf k}$-dependence of the interaction term $2 \beta V({\bf k})$ in the area of the 
splitting is dominant. In this case the profile of the intensity closely follows that 
of the interaction, and there exists one-to-one correspondence between the $V({\bf k})$ 
minima and the $I({\bf k})$ peaks. The ${\bf k}$-dependence of the self-energy in this 
part of the reciprocal space is relatively weak, though qualitatively important for the 
description of the temperature-dependent splitting. Such assumption is certainly 
correct at sufficiently high temperatures, where the KCM approximation (in which the 
self-energy is ${\bf k}$-independent) works reasonably well. Meanwhile, as temperature 
decreases, $\Sigma({\bf k})$ grows faster than $2 \beta V({\bf k})$, since the first 
correction to the KCM self-energy is of order $(\beta V)^2$ (Tsatskis 1998a,b). 
We can then encounter a situation where the variations of $\Sigma({\bf k})$ and 
$2 \beta V({\bf k})$ with the wavevector are comparable. With temperature decreasing 
further, the ${\bf k}$-dependence of the self-energy may even become considerably 
stronger; the $I({\bf k})$ peak positions would then be determined by the 
$\Sigma({\bf k})$ maxima.

\begin{figure}
\begin{center}
\includegraphics[angle=-90]{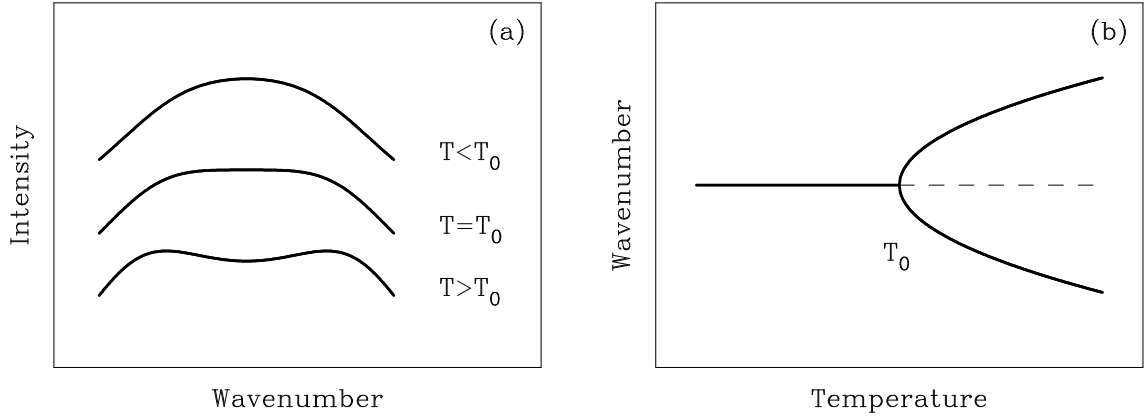}
\end{center}
\caption{Schematic temperature dependence of the SRO intensity profile (a) 
and of the peak positions (b).}
\label{f2}
\end{figure}

The self-energy and the interaction are, in general, qualitatively different functions 
of the wavevector. In particular, there is no special reason to expect $\Sigma({\bf k})$ 
to have any extrema away from the SPs. Therefore, at lower temperatures $I({\bf k})$ 
might no longer exhibit features characteristic for $V({\bf k})$. We assume that 
{\em (i)} $\Sigma({\bf k})$ has extrema only at the SPs, and {\em (ii)} its variation 
with ${\bf k}$ in the area of the splitting becomes more and more important in 
comparison with that of $2 \beta V({\bf k})$ as temperature decreases. Then the 
qualitative picture of the temperature behaviour of the splitting is as follows: At 
high temperatures $\Sigma({\bf k})$ is almost ${\bf k}$-independent, and the 
$I({\bf k})$ peak positions deviate little from the minima of $V({\bf k})$. As 
temperature decreases, the ${\bf k}$-dependence of $\Sigma$ becomes more pronounced; 
the peaks move farther away from the $V({\bf k})$ minima and towards that SP ${\bf k}_0$ 
(either $X$ or $W$) at which $\Sigma({\bf k})$ has a maximum. Eventually, as temperature 
reaches certain value $T_0$, {\em the intensity peaks coalesce at this SP and the 
splitting disappears} (Fig.~\ref{f2}). This effect has never been observed experimentally, 
though Reichert {\em et al.} (1996) assumed that {\em (i)} the coalescence took place 
at the first-order transition temperature $T_t$ (denoted there by $T_0$) and {\em (ii)} 
the splitting grew with increasing temperature as $(T-T_t)^s$; the bifurcation exponent 
$s = 0.38 \pm 0.15$ was determined by fitting to experimental data. The coalescence 
temperature $T_0$ can be found from the condition of vanishing second derivative of 
$I(k)$ at the SP $k_0$, since its sign controls the presence or absence of the splitting. 
At the SPs all the first derivatives vanish, and from Eq.~(\ref{1}) it follows that
\begin{equation}
\left( \partial^2_k I \right)_{k_0} = c(1-c) \, I^2(k_0) 
\left( \partial^2_k \Sigma - 2 \beta \, \partial^2_k V \right)_{k_0} . \label{3}
\end{equation}
The splitting therefore disappears when curvatures of the self-energy and of the 
interaction term at the SP $k_0$ compensate each other: 
\begin{equation}
2 \left( \partial^2_k V \right)_{k_0} = 
T \left( \partial^2_k \Sigma \right)_{k_0} . \label{3a}
\end{equation}

To analyse qualitatively the behaviour of the splitting close to the coalescence 
point, it is convenient to use the analogy with the Landau theory of second-order 
phase transitions (e.g., Landau and Lifshitz 1980). In this temperature range the 
splitting above $T_{0}$ is small, and expansions of $V(k)$ and $\Sigma(k)$ in powers 
of the deviation $\delta k = k - k_0$ of the wavenumber from the SP $k_0$ can be used. 
To describe the split minimum of $V(k)$, only the second- and fourth-order terms are 
needed; since a SP serves as the origin, the expansions do not contain odd powers of 
$\delta k$. It is therefore assumed that in the area of the splitting $V(k)$ and 
$\Sigma(k)$ have the following approximate form, 
\begin{equation}
f(k) = f(0) + a_f (\delta k)^2 / 2 + b_f (\delta k)^4 / 4 \ , \label{4} \\
\end{equation}
where $f=V,\Sigma$, $a_V<0$, $b_V>0$, $a_{\Sigma}<0$ ($\Sigma(k)$ 
has a maximum at $k=k_0$), and the sign of $b_{\Sigma}$ is arbitrary. 
Substituting Eqs.~(\ref{4}) for $V$ and $\Sigma$ into Eq.~(\ref{1}), we 
get the same expression~(\ref{4}) for the inverse intensity $G=I^{-1}$, 
where at $T=T_0$ the second-order coefficient 
$a_G = c(1-c) (-a_{\Sigma} + 2 \beta a_V)$ vanishes (see Eq.~(\ref{3a})), 
while the fourth-order one $b_G$ remains positive. In the vicinity of 
$T_0$ we can then write $a_G = \tilde{a}_G (T-T_0)$, $\tilde{a}_G < 0$ 
and regard $b_G$ as temperature-independent. The inverse intensity 
$I^{-1}(k)$ thus behaves in almost the same way as the Landau free energy, 
and $\delta k$ is the analogue of the order parameter. The only difference 
here is that the role of temperature is reversed: $I^{-1}(k)$ has a double 
minimum above the coalescence temperature $T_0$ and a single one below it. 
From this analogy it follows that the splitting increases with temperature
as $(T-T_0)^{1/2}$ at small positive values of $T-T_0$. Unlike its 
Landau-theory counterpart, the obtained bifurcation exponent $s=0.5$ is 
exact, since the intensity is a regular function of the wavevector 
expandable into the Taylor series. At higher temperatures the behaviour 
of the splitting changes, and its value starts to approach that of the 
splitting in $V(k)$. In this regime the difference between the two 
decreases with temperature as $1/T$, unless the alloy is equiatomic, in 
which case the decrease is faster (Tsatskis 1998a).

Returning to the quantitative description of the temperature dependence of 
the splitting in the whole range of temperatures, we note that it can be 
considerably simplified by taking account of the particular behaviour of $V(k)$ 
and $\Sigma(k)$ (Tsatskis 1998a). The Fourier transform $F(k)$ of an arbitrary 
real-space matrix $F$ defined on the FCC lattice varies with $k$ as 
follows: 1st coordination shell does not contribute to its $k$-dependence; 2nd 
and higher shells lead to the term proportional to $\cos 2 \pi k$; starting from 
8th shell, $\cos 4 \pi k$ term appears; 21st shell produces $\cos 6 \pi k$ term, 
etc. Here $k$ is defined as the deviation from the SP X and measured in the 
reciprocal-lattice units (r.l.u.). As a result, under the assumption that matrix 
elements of $F$ beyond 20th coordination shell are of no importance, 
\begin{equation}
F(k) = A_F + 2 B_F \cos 2 \pi k + 2 C_F \cos 4 \pi k \ , \label{7}
\end{equation}
where the coefficients are linear combinations of the matrix elements $F_{ij}$; 
for explicit expressions for the relevant quantities $B_F$ and $C_F$, see 
Tsatskis (1998a). To find the $I(k)$ peak positions, we insert Eqs.~(\ref{7}) 
for $V$ and $\Sigma$ into Eq.~(\ref{2}) for the $I(k)$ extrema. The presence 
of the off-SP minima in $V(k)$ implies $C_{V}>0$, $|B_{V}| < 4 C_{V}$, while 
the assumption of the absence of such extrema in $\Sigma(k)$ requires 
$|B_{\Sigma}| > 4 |C_{\Sigma}|$. The positions $k_I$ of the intensity peaks 
away from the SPs above $T_0$ are given by 
\begin{equation}
\cos 2 \pi k_I = - \left( 2 B_V - T B_{\Sigma} \right) / 
4 \left( 2 C_V - T C_{\Sigma} \right) \ , \label{9}
\end{equation}
whereas the $V(k)$ minima $k_V$ can be obtained from Eq.~(\ref{9}) by putting 
$B_{\Sigma}=C_{\Sigma}=0$. Substituting the same Eqs.~(\ref{7}) into Eq.~(\ref{3a}), 
we obtain a simpler equation for the coalescence temperature $T_0$,
\begin{equation}
2 \left( \pm B_V + 4 C_V \right) = T_0 \left( \pm B_{\Sigma} 
+ 4 C_{\Sigma} \right)_{T_0} \ , \label{11}
\end{equation}
where upper and lower signs correspond to the coalescence at the SPs X and W, 
respectively. Contrary to Eq.~(\ref{9}), the explicit expression for $T_0$ cannot 
be obtained, since $B_{\Sigma}$ and $C_{\Sigma}$ are temperature-dependent. 

\begin{figure}
\begin{center}
\includegraphics[angle=-90]{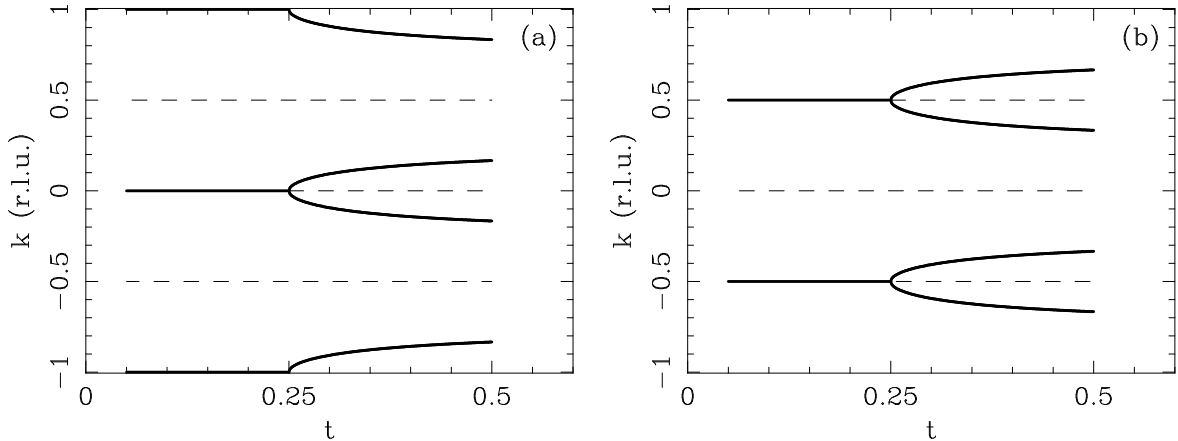}
\end{center}
\caption{Positions of the intensity peaks versus dimensionless temperature 
$t = 4 C_V T / |B_W|$ for the A$_3$B alloy with the interaction $V({\bf k})$ such 
that $C_W = 0$ and $k_V = 0.25$ (i.e., the minima of $V(k)$ are located half-way 
between SPs $X$ and $W$). For this set of parameters $t_0^X = t_0^W = 0.25$, and 
the character of the temperature dependence of the splitting is defined by the sign 
of $B_W$: (a) $B_W > 0$, case {\em (i)} in the text, (b) $B_W < 0$, case {\em (ii)}.}
\label{f3}
\end{figure}

To calculate $B_{\Sigma}$ and $C_{\Sigma}$, a particular approximation for SRO must 
be used. Here we turn to the simplest theory leading to the wavevector dependence 
of the self-energy, the high-temperature expansion (HTE) (Tsatskis 1998a,b). 
The HTE is the expansion in powers of $\beta V$; the second-order HTE approximation 
for the self-energy gives
\begin{equation}
\Sigma({\bf k}) = \Sigma_{d} + 2 x \beta^{2} W({\bf k}) \ , \label{12}
\end{equation}
where $x=(1-2c)^{2}$, $W_{ij}=V_{ij}^{2}$, and $\Sigma_{d}$ is the diagonal part 
of the self-energy in the site representation which does not contribute to the 
${\bf k}$-dependence of $\Sigma$. Then 
\begin{equation}
B_{\Sigma} = 2 x \beta^{2} B_W \ , \ \ \ 
C_{\Sigma} = 2 x \beta^{2} C_W \ , \label{13}
\end{equation}
where $|B_W| > 4 |C_W|$. Substitution of Eqs.~(\ref{13}) into Eq.~(\ref{9}) 
leads to the result
\begin{equation}
\cos 2 \pi k_I = - \left( B_V - x \beta B_W \right) / 
4 \left( C_V - x \beta C_W \right) \ . \label{14}
\end{equation}
Calculation of the temperature derivative shows that in this approximation the 
behaviour of the splitting is determined by the sign of the $T$-independent quantity 
$B_{W} C_{V} - B_{V} C_{W}$:
\begin{equation}
\partial_{\beta} \left( \cos 2 \pi k_I \right) = 
x \left( B_{W} C_{V} - B_{V} C_{W} \right) / 
4 \left( C_{V} - x \beta C_{W} \right)^{2} \ . \label{14a}
\end{equation}
{\em (i)} $B_{W} C_{V} > B_{V} C_{W}$: The derivative~(\ref{14a}) is positive, 
and $\cos 2 \pi k_I$ increases with decreasing temperature; the $I({\bf k})$ peaks 
move away from the $V({\bf k})$ minima and towards the SP X (Fig.~\ref{f3}(a)). 
Finally, the peaks coalesce at X when $\cos 2 \pi k_I$ reaches the value +1 at 
the bifurcation point
\begin{equation}
T_0^X = x \left( B_W + 4 C_W \right) / \left( B_V + 4 C_V \right) \ . \label{15a}
\end{equation}
{\em (ii)} $B_{W} C_{V} < B_{V} C_{W}$: The intensity peaks shift towards the 
SP W (Fig.~\ref{f3}(b)). The splitting disappears when $\cos 2 \pi k_I = -1$, i.e., 
at the temperature
\begin{equation}
T_0^W = x \left( B_W - 4 C_W \right) / \left( B_V - 4 C_V \right) \ . \label{15b}
\end{equation}
{\em (iii)} $B_{W} C_{V} = B_{V} C_{W}$, or the alloy is equiatomic ($c=0.5$): The 
derivative~(\ref{14a}) vanishes and the second-order HTE approximation predicts 
temperature-independent splitting. In this situation the temperature behaviour of 
the splitting is defined by the third- and higher-order terms in the HTE for the 
self-energy. 

The predicted coalescence of the FS-related intensity peaks can, in principle, be 
described by using any theory of SRO which leads to the ${\bf k}$-dependent self-energy. 
As was already noted, the frequently used KCM and SM approximations fail to reproduce 
this feature of the exact self-energy. Apart from the HTE, the available theories include 
the cluster variation method (CVM) (Kikuchi 1950, 1951), the $\gamma$-expansion method 
(GEM) (Tokar 1985, Tokar, Masanskii and Grishchenko 1990, Masanskii, Tokar, and 
Grishchenko 1991), and the $\alpha$-expansion (AE) (Tsatskis 1998a) closely related to 
the GEM. The corresponding approximations for the self-energy were compared by Tsatskis 
(1998b). However, the CVM is a direct-space method and as such is hardly applicable to 
the case of long-range interactions. To describe the off-SP minima of $V({\bf k})$, at 
least eight first potentials are needed, and accurate sets of inverse Monte Carlo (MC) 
interactions determined from experimental data are yet significantly larger 
(Sch\"{o}nfeld {\em et al.} 1996). The HTE approach is limited to the same extent as 
the KCM approximation; it leads to quantitatively correct results only at sufficiently 
high temperatures, as follows from its very name. The GEM utilizes the assumption 
about rapid decrease of interatomic correlations with distance, which is invalid for 
long-range interactions. This leaves the AE, a modification of the GEM specifically 
designed to be used in such cases (Tsatskis 1998a). Similarly to the GEM, the AE is 
expected to be sufficiently accurate at realistic temperatures. 

We now compare the effect of the coalescence with another anomaly in the SRO 
scattering, the ``thermal'' splitting of the intensity peak, observed for the Pt-V 
alloy system and reproduced in the MC simulations (Le Bolloc'h, Cren, Caudron and 
Finel 1997). For this system, the (100) peak splits along the (h00) line with increasing 
Pt content. A theory of this anomaly was proposed by Tsatskis (1998c), who also showed 
that the splitting can occur when temperature decreases at fixed composition. This 
prediction was confirmed by the MC calculations (Le Bolloc'h 1997), though the effect 
has not yet been observed experimentally. According to the proposed theory, the thermal 
splitting can be explained by the same mechanism of the compensation at the SP of the 
$\Sigma({\bf k})$ and $2 \beta V({\bf k})$ curvatures. However, in this case the 
interaction $V({\bf k})$ has a simple minimum at the corresponding SP and, in 
agreement with the mean-field arguments, produces the $I({\bf k})$ peak with no 
fine structure at higher temperatures. This peak splits as temperature decreases and 
the {\em positive} curvature of $\Sigma({\bf k})$ exceeds that of $2 \beta V({\bf k})$. 
In the event of the coalescence, on the other hand, the fine structure of $V({\bf k})$ 
is erased at lower temperatures due to the growing {\em negative} $\Sigma({\bf k})$ 
curvature. Both phenomena can be conveniently described within the same Landau-type 
approach outlined above. The only formal difference is the sign of $\tilde{a}_G$: 
$\tilde{a}_G > 0$ for the thermal splitting and $\tilde{a}_G < 0$ for the coalescence. 

The present theory shows that the coalescence temperature $T_0$ does not, in general, 
coincide with the transition temperature $T_t$ and is more or less arbitrary. 
Nevertheless, it can accidentally be close to $T_t$ (unless the ordering transition 
occurs first). This is a possible reason for the rather good agreement between the 
theoretical value $s=0.5$ of the bifurcation exponent and the experimental result 
$s = 0.38 \pm 0.15$ obtained for the Cu$_{3}$Au alloy (Reichert {\em et al.} 1996). 
The remaining discrepancy can be attributed to the difference between $T_0$ and $T_t$, 
and also to the data fitting over the wide temperature interval (about 140 K).

Finally, we note that the proposed mechanism of the curvature compensation explains 
also the results obtained for the exactly solvable 1D Ising model with nearest- and 
next-nearest-neighbour interactions (Kulik, Gratias and de Fontaine 1989) and for the 
2D ANNNI model in the framework of the CVM (Finel and de Fontaine 1986). Observed for 
these two models is, in fact, the same effect of the coalescence as described here: 
the $I({\bf k})$ peaks produced by the off-SP minima of $V({\bf k})$ move towards a 
SP as temperature decreases and transform at this SP into a single peak at some 
temperature before the onset of long-range order (for the 1D model, at non-zero 
temperature). The exact solution in the former and the CVM in the latter case lead 
to the ${\bf k}$-dependent self-energy whose SP curvature compensates that of the 
interaction term at the coalescence temperature. In all three cases the real-space 
interactions produce minima off SPs in the reciprocal space; however, Finel and de 
Fontaine (1986) and Kulik {\em et al.} (1989) considered the short-range competing 
interactions which are very different from the long-range interactions associated with 
the FS effects.

In summary, we have pointed out a possibility for the FS-related splitting of the SRO 
intensity peak to disappear at some temperature above the order-disorder transition. The 
off-SP intensity peaks reflecting corresponding minima of the effective pair interaction 
in the ${\bf k}$-space can move towards one of the SPs (X or W) as temperature decreases 
and eventually coalesce at this SP, transforming themselves into a single peak. The 
driving force behind this effect is the ${\bf k}$-dependence of the self-energy which 
becomes stronger with decreasing temperature. The coalescence occurs when the self-energy 
and interaction curvatures at the SP compensate each other. The shift of the intensity 
peaks with decreasing temperature towards X (an increase of the splitting with 
temperature) was observed for the Cu$_{3}$Au alloy (Reichert {\em et al.} 1996), while 
neither the shift towards W (a decrease of the splitting with temperature) predicted by 
Tsatskis (1998a) nor the coalescence at any of these SPs were found. The coalescence can 
be described qualitatively by the Landau-type theory and quantitatively by any theory of 
SRO which is sufficiently accurate and leads to the wavevector dependence of the 
self-energy. This excludes two popular approximations, the KCM and the SM, in which the 
self-energy is site-diagonal. The effect of the coalescence has been compared with the 
``thermal'' splitting of the SRO intensity peak which was observed experimentally and in 
the MC simulations and can be described within the same theoretical framework.

\pagebreak

\begin{center}
{\bf REFERENCES} \\ 
\end{center}
$^\dag$Electronic address: it10001@cus.cam.ac.uk \\
$^\ddag$Former name: I. V. Masanskii \\ 
Brout, R., 1965, {\em Phase Transitions} (New York: Benjamin). \\ 
Clapp, P.C., and Moss, S.C., 1966, Phys. Rev. {\bf 142}, 418; 1968, {\bf 171}, 754. \\ 
Finel, A., and de Fontaine, D., 1986, J. Statist. Phys. {\bf 43}, 645. \\ 
Kikuchi, R., Phys. Rev., 1950, {\bf 79}, 718; 1951, {\bf 81}, 988. \\ 
Krivoglaz, M.A., 1969, {\em Theory of X-Ray and Thermal Neutron 
Scattering by Real Crystals} (New York: Plenum); \\ 
\hspace*{10mm} 1996, {\em Diffuse Scattering of X-Rays and Neutrons 
by Fluctuations} (Berlin: Springer). \\ 
Kulik, J., Gratias, D., and de Fontaine, D., 1989, Phys. Rev. B {\bf 40}, 8607. \\ 
Landau, L.D., and Lifshitz, E.M., 1980, {\em Statistical Physics}, Part~1 
(Oxford: Pergamon). \\ 
Le Bolloc'h, D., 1997, private communication. \\ 
Le Bolloc'h, D., Cren, T., Caudron, R., and Finel, A., 1997, 
Comput. Mater. Sci. {\bf 8}, 24. \\ 
Masanskii, I.V., Tokar, V.I., and Grishchenko, T.A., 1991, Phys. Rev. B {\bf 44}, 4647. \\ 
Ohshima, K., and Watanabe, D., 1973, Acta Cryst. A {\bf 29}, 520. \\ 
Ohshima, K., and Watanabe, D., and Harada, J., 1976, Acta Cryst. A {\bf 32}, 883. \\ 
Reichert, H., Moss, S.C., and Liang, K.S., 1996, Phys. Rev. Lett. {\bf 77}, 4382. \\ 
Reichert, H., Tsatskis, I., and Moss, S.C., 1997, Comput. Mater. Sci. {\bf 8}, 46. \\ 
Saha, D.K., Koga, K., and Ohshima, K., 1992, J. Phys.: Condens. Matter {\bf 4}, 10093. \\ 
Saha, D.K., and Ohshima, K., 1993, J. Phys.: Condens. Matter {\bf 5}, 4099. \\ 
Scattergood, R.O., Moss, S.C., and Bever, M.B., 1970, Acta Metall. {\bf 18}, 1087. \\ 
Sch\"{o}nfeld, B., Roelofs, H., Malik, A., Kostorz, G., Plessing, J., and Neuhauser, H., 
1996, Acta Mater., {\bf 44}, 335. \\ 
Tokar, V.I., 1985, Phys. Lett. A {\bf 110}, 453. \\ 
Tokar, V.I., Masanskii, I.V., and Grishchenko, T.A., 1990, 
J. Phys.: Condens. Matter {\bf 2}, 10199. \\ 
Tsatskis I., 1998a, J. Phys.: Condens. Matter {\bf 10}, L145, 
preprint cond-mat/9801089; 1998b, in {\em Local Structure from \\ 
\hspace*{10mm} Diffraction}, Fundamental Materials Science Series, 
eds. M.F. Thorpe and S.J.L. Billinge (New York: Plenum \\ 
\hspace*{10mm} Press), preprint cond-mat/9803052; 
1998c, to be published in Phys. Lett. A, preprint 
cond-mat/9801090. 

\end{document}